\documentclass{emulateapj}
\usepackage{times}

\newcommand{\beq}{\begin{equation}}
\newcommand{\eeq}{\end{equation}}

\def\alp{\mbox{$\alpha$}}

\def\arcmin{\hbox{$^\prime$}}
\def\arcsec{\hbox{$^{\prime\prime}$}}
\def\solar{\mbox{$_{\normalsize\odot}$}}

\def\deg{\hbox{$^\circ$}}

\newcommand{\AmS}{{\protect\the\textfont2
  A\kern-.1667em\lower.5ex\hbox{M}\kern-.125emS}}
\newcommand{\lsim}{\ \raise
-2.truept\hbox{\rlap{\hbox{$\sim$}}\raise5.truept\hbox{$<$}\ }}
\newcommand{\gsim}{\ \raise
-2.truept\hbox{\rlap{\hbox{$\sim$}}\raise5.truept\hbox{$>$}\ }}
\newcommand{\simsim}{\ \raise
-2.truept\hbox{\rlap{\hbox{$\sim$}}\raise5.truept\hbox{$\sim$}\ }}

\slugcomment{Accepted for Publication in the Astrophysical Journal}

\shorttitle{Triggered Star Formation in NGC~346/N~66 in the SMC}
\shortauthors{D. A. Gouliermis et al.}

\begin{document}

\title{NGC~346 in The Small Magellanic Cloud. IV.  Triggered Star
Formation in the {\sc H~ii} Region N~66}





\author{Dimitrios A. Gouliermis\altaffilmark{1}, 
        You-Hua Chu\altaffilmark{2}, 
        Thomas Henning\altaffilmark{1},\\
        Wolfgang Brandner\altaffilmark{1}, 
        Robert A. Gruendl\altaffilmark{2}, 
        Eva Hennekemper\altaffilmark{1}, and 
        Felix Hormuth\altaffilmark{1}
}

\email{dgoulier@mpia.de, chu@astro.uiuc.edu, henning@mpia.de,
brandner@mpia.de, gruendl@astro.uiuc.edu, hennekemper@mpia.de,
hormuth@mpia.de}

\altaffiltext{1}{Max Planck Institute for Astronomy, K\"onigstuhl 17,
69117 Heidelberg, Germany}
\altaffiltext{2}{Department of Astronomy, University of Illinois, 1002
West Green Street, Urbana, IL 61801}


\begin{abstract}

Stellar feedback, expanding {\sc H~ii} regions, wind-blown bubbles, and
supernovae are thought to be important triggering mechanisms of star
formation.  Stellar associations, being hosts of significant numbers of
early-type stars, are the loci where these mechanisms act. In this part
of our photometric study of the star-forming region NGC~346/N~66 in the
Small Magellanic Cloud, we present evidence based on previous and recent
detailed studies, that it hosts at least two different events of
triggered star formation and we reveal the complexity of its recent star
formation history.  In our earlier studies of this region (Papers I,
III) we find that besides the central part of N~66, where the bright OB
stellar content of the association NGC~346 is concentrated, an arc-like
nebular feature, north of the association, hosts recent star formation.
This feature is characterized by a high concentration of emission-line
stars and Young Stellar Objects, as well as embedded sources seen as
IR-emission peaks that coincide with young compact clusters of low-mass
pre-main sequence stars. All these objects indicate that the northern
arc of N~66 encompasses the most current star formation event in the
region. We present evidence that this star formation is the product of a
different mechanism than that in the general area of the association,
and that it is triggered by a wind-driven expanding {\sc H~ii} region
(or bubble) blown by a massive supernova progenitor, and possibly other
bright stars, a few Myr ago. We propose a scenario according to which
this mechanism triggered star formation away from the bar of N~66, while
in the bar of N~66 star formation is introduced by the photo-ionizing OB
stars of the association itself.

\end{abstract}

\keywords{Magellanic Clouds --- stars: winds, outflows --- ISM: bubbles 
--- HII regions --- supernova remnants --- clusters: individual (NGC~346)}

\section{Introduction}

Massive OB stars, not having an optically visible pre-main sequence
(PMS) contraction phase, appear almost immediately after their birth on
the main sequence (e.g. Stahler \& Palla 2005). They are mostly grouped
in stellar associations, loose concentrations of stars, which host also
significant numbers of intermediate- and low-mass PMS stars (see review
by Brice\~{n}o et al. 2007). When the far-UV radiation of the bright OB
stars reaches the surface of the parental molecular cloud, a
photo-dissociated region (PDR) develops. Nebula LHA~115-N~66 or in short
N~66 (Henize 1956), the brightest {\sc H~ii} region in the Small
Magellanic Cloud (SMC), being very rich in early-type stars, is
certainly an excellent example of an extragalactic PDR. The stellar
association NGC~346, located in the central part of the nebula, hosts
the largest sample of spectroscopically confirmed OB stars in the SMC
(Massey et al. 1989), which have been the subject of several previous
investigations (Niemela et al.  1986; Walborn et al. 2000; Evans et al. 
2006).

These massive stars evolve rapidly, and immediately start to ionize the
cloud, blowing its material away. The degree of ionization decreases
outwards, and a thin barrier develops that segregates the ionized from
the atomic gas. Molecular hydrogen is not fully ionized behind this
front, but partly dissociated, while CO molecules which are located a
bit deeper into the cloud are more easily dissociated than H$_{2}$ by
absorbing UV photons. A correlation between H$_{2}$ infrared emission
and CO lines, characteristic of a PDR, has been found for the region of
NGC~346/N~66 by Contursi et al. (2000) and Rubio et al. (2000). These
authors define the ``bar'' of N~66 as the oblique bright emission
region, extending from southeast to northwest centered on NGC~346. They
suggest that {\sl star formation has taken place as a sequential process
in the bar of N~66}, resulting in several embedded sources, seen as
IR-emission peaks in the 2.14~\micron\ H$_{2}$ line and the ISOCAM LW2
band (5 - 8 \micron). These peaks are alphabetically numbered from ``A''
to ``I'' (Contursi et al. 2000), with the association NGC~346 itself
coinciding with peak ``C''.

Recent studies reconstruct the star formation history in nearby Galactic
OB associations, and provide observational evidence for sequential and
triggered star formation in their vicinity. Preibisch \& Zinnecker
(2007) conclude from their study of the Scorpius-Centaurus OB
association that the formation of whole OB subgroups (each consisting of
several thousand stars) requires large-scale triggering mechanisms such
as shocks from expanding wind and supernova driven super-bubbles
surrounding older subgroups\footnote{According to these authors, other
triggering mechanisms, like radiatively driven implosion of globules,
also operate, but seem to be secondary processes, forming only small
stellar groups rather than whole OB subgroups with thousands of stars.}.
Since the low-mass stellar members of associations remain in their PMS
phase for \lsim~30~Myr, they play a key r\^{o}le in the understanding of
star formation in the vicinity of these systems (Brice\~{n}o et al.
2007). Consequently, the recent discovery of a plethora of low-mass PMS
stars in the region of NGC~346/N~66 with photometry from the {\sl 
Advanced Camera for Surveys} (ACS) onboard HST (Nota et al. 2006;
Gouliermis et al. 2006, hereafter Paper~I) can contribute significantly
to the clarification of the mechanisms that may act in this
extraordinary star-forming region.

The subsequent investigation of these PMS stars in NGC~346/N~66 (Sabbi et 
al. 2007, hereafter SSN07; Hennekemper et al. 2008, hereafter Paper~III) 
demonstrated that indeed they are clustered in several compact 
concentrations, some of them coinciding with the IR-emission peaks of 
Contursi et al. (2000; detected also with {\sl Spitzer}, see \S 2.1), 
verifying the existence of stellar subgroups in the region, similarly to 
galactic associations. Not being able to resolve differences in age 
smaller than 1 - 2 Myr in the individual CMDs of these sub-clusters, SSN07 
suggest that ``all sub-clusters appear coeval with each other''. According 
to these authors, this coevality is a signature of the star formation 
conditions predicted by the hierarchical fragmentation of a turbulent 
molecular cloud model (Klessen \& Burkert 2000; Bonnell \& Bate 2002; 
Bonnell et al. 2003).

However, our analysis on the clustering properties of the PMS stars in
NGC~346/N~66 (Paper~III) showed that {\sl there are significant age
differences between some sub-clusters and the association itself}. 
Specifically, three compact PMS clusters located to the north of the bar
of N~66 are found to be not older than 2.5~Myr, while the CMD of the
main body of the association NGC~346 show indications of an underlying
older PMS population. This clearly suggests that these northern clusters
are probably formed {\sl after} the central stellar association. As we
discuss later, triggering mechanisms for star formation such as those
described by models of ionization shock fronts from OB stars (e.g.
Kessel-Deynet \& Burkert 2003) or wind-driven shock waves (e.g. Vanhala
\& Cameron 1998) may explain better their formation. Indeed, while a
sequential star formation mechanism has been suggested to take place
{\sl in} the bar of N~66, around the association (Rubio et al. 2000),
the existence of young compact PMS clusters {\sl away} from it to the
north of the association, does not quite fit in the hypothesis that this
mechanism propagated from the center of N~66 along the bar. The
triggering agent of these clusters may well be located outside of the
bar.

In this paper we consider the findings from earlier and recent
comprehensive investigations of the region NGC~346/N~66 to attempt a
clearer understanding of the mechanisms that shape the recent star
formation of this outstanding extragalactic star-forming region. Our aim
is to provide answers to two important questions: i) {\sl Is the star
formation away from the association NGC~346 the product of triggered
fragmentation of the cloud alone?} ii) {\sl Is the photo-dissociation by
the early-type stars of NGC~346 in the center of N~66 the only
triggering mechanism in the region?} In section~2 we present evidence
that this region has a far more complicated recent star formation
history than what was previously considered, and that the most recent
star formation event could not have been triggered by the central
association. In section~3 we propose a scenario for the recent star
formation in the northern part of the region NGC~346/N~66 and we
indicate a nearby massive supernova progenitor, located at the northeast
of the bar of N~66, as the triggering agent. We also discuss the
possibility that other nearby massive stars away from the association
may have also contributed to this mechanism. We present supporting
evidence to this scenario using analyses of massive stellar content and
gas kinematics of this region. Concluding remarks are given in
section~4.


\section{Star Formation in the region of NGC~346/N~66}

\subsection{Star Formation away from the bar of N~66}

Recent {\sl Spitzer} IR observations of NGC~346/N~66 revealed a rich
sample of embedded sources, identified as candidate Young Stellar
Objects (YSOs; Bolatto et al. 2007; Simon et al. 2007). A large
concentration of these sources are located in the bar of N~66, in
agreement with the distribution of the PMS stars found with our ACS
photometry (Paper~III), and they coincide with the emission peaks of
this area identified by Rubio et al. (2000; peaks A, B, C, D, E, H and
I; see also their Fig. 8).  However, another important concentration of
YSOs is located to the north of the bar, coinciding with two additional
emission peaks (peaks F and G) and with a second concentration of PMS
stars, which forms an arc-like feature, extending from southwest to
northeast. The ACS image of the region in H\alp\ (Paper~III), the
8\micron\ {\sl Spitzer} image (Simon et al. 2007), as well as the maps
of the CO(2$-$1) line emission (Rubio et al. 2000) give further support
for the existence of this feature\footnote{The description of this
feature as an ``arc-like'' is based on the gas emission in the region
seen in H\alp\ and [{\sc O~iii}] mostly from its southern part. In CO
and 8\micron-emission, which dominates its northern part the morphology
of this feature is more linear, but still curved, and closer to
``arm-like''.}. Three compact clusters of PMS stars are identified by us
(Paper~III) and SSN07 in this arc-like feature (clusters 2a, 2b, and 3
in Paper~III, and Sc-13, 14 and 15 in SSN07) to coincide with both
IR-emission peaks. The PMS stars in these clusters are found to be
younger than those of the bar, implying that they are more recently
formed. Moreover, this northern feature is also characterized by a high
concentration of emission-line stars found with our ACS photometry in
H{\alp} (Paper~III). These facts clearly suggest that indeed star
formation currently takes place in an extended feature away from the bar
of N~66, northeast of NGC~346.

\begin{figure*}[t!]
\epsscale{1.15}
\plottwo{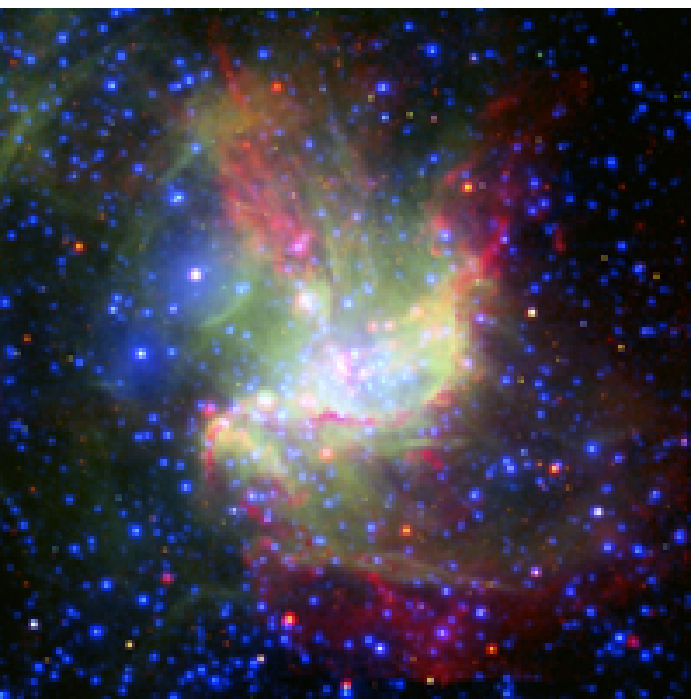}{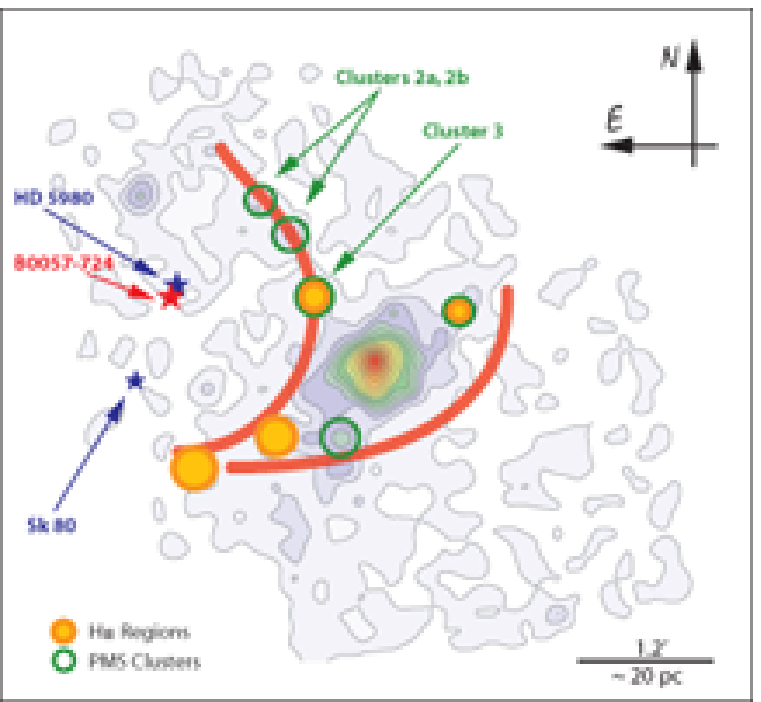}
\caption{{\sl Left}: Color-composite image of the area of NGC~346/N~66
from combination of X-ray observations with XMM-Newton (blue), [{\sc O
iii}] (501.1~nm) with NTT (green) and at 8\micron\ with {\sl Spitzer}
(red). The image covers an area about 6\arcmin~$\times$~6\arcmin\ ($\sim
100 \times 100$~pc$^2$) wide centered on NGC~346. North is up, East is
to the left. {\sl Right}: Schematic representation of the proposed star
formation scenario, according to which the massive progenitor of
SNR~B0057$-$724 triggered current star formation at the north of the
N~66 bar through its wind-blown bubble.  The contribution of the most
massive stellar object in the area, the LBV HD~5980, is also considered
important. The colored iso-density contour map, constructed from star
counts of both bright main sequence and faint PMS stars of the region,
is shown in the background to demonstrate the geometry of the youngest
stellar concentrations. The locations of the SNR, HD~5980, the O-star
Sk~80 and significant PMS clusters and {\sc H~ii} regions are also
indicated. Two important dusty features, also visible in the left panel,
are drawn with red thick curves. The one to the south, south-west of the
bar of N~66 is considered to be the product of the energy output from
the OB stars of the association NGC~346. The other to the north and east
of the bar coincides very well with a shock-wave coming from the
direction of the SNR~B0057$-$724.  The synergy of both these strong
energy sources seems to define the recent star formation history of the
region. \label{fig-imaxoi}}
\end{figure*}

This feature can be easily distinguished in the color-composite image 
shown in the left panel of Fig.~\ref{fig-imaxoi}. We constructed this 
image from observations of the general area around NGC~346/N~66 with 
XMM-Newton in X-rays (blue), ESO NTT in [{\sc O~iii}] (green) and {\sl 
Spitzer}/IRAC in the 8\micron\ band (red). The XMM-Newton observations 
have ID 0110000201 (PI: J. Bleeker), and they have been used in 
variability studies of X-ray sources (Naz\'{e} et al. 2004) and in the 
investigation of high mass X-ray binaries (Sasaki, Pietsch \& Haberl 2003; 
Shtykovskiy \& Gilfanov 2005) in the region of NGC~346/N~66. The [{\sc 
O~iii}] (501.1 nm) image of NGC~346/N~66 was taken with NTT/EMMI within 
the ESO Program 56.C-0379 (PI: I. J. Danziger) and it was also presented 
in the analysis by Rubio et al. (2000).  The 8\micron\ imaging data have 
been obtained with the Infrared Array Camera (IRAC; Fazio et al. 2004) 
onboard the {\sl Spitzer Space Telescope} within the GTO Science Program 
63 (PI: J. R. Houck) on 2004 April 20, and it has been used for the 
detection of candidate YSOs in the region (Simon et al. 2007).

The association NGC~346 is located within the bar of N~66, almost at the
center of the image of Fig.~\ref{fig-imaxoi} (left panel). From this
image, as well as from the ones taken with {\sl HST}/ACS (see Nota et
al. 2006) and {\sl Spitzer}/IRAC (see Simon et al. 2007) it can be seen
that there is a relation between NGC~346 and a southern dusty arc
feature. This one seems to outline the ionization front of the remaining
cloud, due to the powerful winds of the OB stars of the association.
However, the shape and orientation of the star-forming feature away and to
the north of NGC~346, located at the top left part of the image,
suggests that it is triggered probably by the same kind of mechanism,
but this process {\sl could not be produced} by the photo-dissociation
of the cloud by the ionizing stars of the association.  If it was so,
this feature would not face to the east, but rather to the south in a
symmetrical manner to the southern arc. Since, this is not the case, the
source that triggered star formation in the northern extended feature
{\sl should not be the central association, but it should be rather
located to the east of the field}.


\subsection{A Multi-wavelength Image of NGC~346/N~66}

Two X-ray bright features can be seen (in blue) in Fig.~\ref{fig-imaxoi}
({\sl left}), located to the east and east-northeast of NGC~346
respectively (left part of the image). In this image an extraordinarily
good agreement between the northern (upper) X-ray nebula (in blue) and
gas presumably photo-ionized seen in [{\sc O~iii}] (in green) is
present. Although the X-ray emission region appears in
Fig.~\ref{fig-imaxoi} to be small and almost circular, it is actually
much more extended, as {\sl Chandra} ACIS observations show (Guerrero \&
Chu 2008; their Fig. 1c), with its western boundary reaching the
northern edge of the bar of N~66, where [{\sc O~iii}] emission provides
more evidence of heated gas. The northern arc-like nebular feature (seen
in green-red) is perfectly outlined by YSOs, IR peaks, and young PMS
stars (discussed in \S~2.1), and the 8\micron\ mid-infrared component of
the color image (shown in red) gives further information about dust
emission from the star-forming arm.  Specifically, its northern part
(which includes two PMS clusters and one IR-emission peak), suggests
that there is probably still undisrupted cloud material in this part of
the region (this is verified by the CO map by Rubio et al. 2000). The
location of this part of the star-forming feature as seen in
Fig.~\ref{fig-imaxoi} ({\sl left}) is not compatible with a sequential
star formation in the bar of N~66, but rather with a mechanism outside
the bar. In order to clarify how this feature may have been actually
formed we discuss in the following sections the two most important
energy sources known in this specific area.

\subsection{HD~5980: A Peculiar Massive Object in N~66}

Embedded in the northernmost diffuse X-ray emission of
Fig.~\ref{fig-imaxoi} is a point source centered on the remarkable
massive triple system HD~5980, located at 00$^{\rm h}$59$^{\rm
m}$26.55$^{\rm s}$, $-$72\deg09\arcmin53.8\arcsec\ (J2000).  This
object, which underwent a luminous blue variable (LBV) type
eruption\footnote{LBVs are thought to be evolved He-burning stars which
may develop into Wolf-Rayet stars.} in 1994 is also known as Sk 78
(Sanduleak 1968)\footnote{The southernmost X-ray bright source is known
as Sk 80, and it is a bright massive star classified as O7Iaf (see also
Evans et al. 2006).}. Its spectral variations are unique: Before 1980
this object was classified as WN+OB, then it changed to WN3+WN4 for the
period 1980 - 1983, and to WN6 in 1992.  In 1994, after the
LBV-eruption, it turned to WN11. The high X-ray emission of HD 5980
makes it comparable to the X-ray-brightest single WN stars (Wessolowski
1996) and the brightest WR+OB binaries (Pollock 2002) of the Galaxy. The
nature of HD~5980 is investigated in detail with {\sl Chandra}
observations of NGC~346 by Naz\'{e} et al. (2002), who note that the
fast wind from the post-eruptive phase, which collides with the slow
wind ejected during the eruption, causes the high X-ray surface
brightness. Bright diffuse X-ray emission has been observed in other
LBVs, i.e. $\eta$ Carinae in the Milky Way. However, in contrast to
$\eta$ Carinae, HD~5980 has not had time to develop a LBV nebula around
it yet, and so another proposed explanation for the high X-ray
luminosity of HD~5980 is the collision of the winds from its close
massive stellar components.

Naturally, this massive object may be associated, due to its energetic 
nature, with the northern feature in the region of NGC~346/N~66, 
and indeed, Walborn (1978) notes that the arcs seen in H$\alpha$ and [{\sc 
O~iii}] are an indication that HD~5980 interacts with its environment. 
However, from comparisons between X-ray and visible data, Naz\'{e} et al. 
(2002) argue that {\sl the bright, extended X-ray emission around HD 5980 
is most probably due to a nearby core-collapse supernova remnant (SNR) 
whose progenitor is unknown}. These authors describe the shape of this 
emission as rectangular with an extension to the northeast.  Its 
brightness is rather uniform with the exception of some dark arcs. Its 
size is $\simeq$ 37~$\times$~29~pc. They also note that the spatial 
coincidence of this extended X-ray emission with the peculiar massive star 
implies an association between the two objects. We explore this hypothesis 
in the following section, where the known information about this SNR is 
presented.


\subsection{A Core-Collapse Supernova Remnant in N~66}

Massive stars in stellar associations affect their environment by ionizing 
radiation, stellar winds, and, finally, supernova explosions (e.g. Chu 
1997). Massive supernova progenitors do not migrate far from their 
birthplaces in their short lifetimes, and when they explode the produced 
shock-waves can trigger cloud collapse, enhancing star formation in the 
vicinity of their environment. Numerical studies (e.g. Vanhala \& Cameron 
1998; Fukuda \& Hanawa 2000) show that shock-waves with velocities in the 
range of 15~-~45~km~s$^{-1}$, corresponding to supernova explosions at a 
distance between 10 pc and 100 pc from the molecular cloud, can induce 
such a collapse.

Earlier studies show that the general area of NGC~346/N~66 is 
characterized by at least three known SNRs (Ye et al. 1991; Reid et al. 
2006)\footnote{Reid et al.  (2006) suggest the existence of another SNR 
candidate in the region at 00$^{\rm h}$59$^{\rm m}$32.4$^{\rm s}$, 
$-$72\deg08\arcmin42.4\arcsec (J2000), based on observations from the DBS 
2.3-m telescope at the Siding Spring Observatory.} located away from the 
center of the region: SNR B0056$-$724 (also known as SNR 0056$-$72.4), 
located to the southwest of the region, SNR B0056$-$725 (or SNR 
0056$-$72.5) to the west, and SNR B0057$-$724 (or SNR 0057$-$72.2) to the 
east. The first two SNRs are located to the south and west, outside the 
field-of-view shown in Fig.  \ref{fig-imaxoi}. The third SNR 
(B0057$-$724), though, is located closer to the central part of 
NGC~346/N~66, and actually {\sl is the SNR associated with the 
northernmost extended X-ray emission} seen in Fig.~\ref{fig-imaxoi} ({\sl 
left}). Indeed, Reid et al. (2006) demonstrated that the extended X-ray 
emission surrounding HD~5980 coincides exactly with the radio non-thermal 
emission, confirming ``beyond a doubt its true nature as a SNR''.

SNR~B0057$-$724 has been previously studied in H$\alpha$ with echelle
spectroscopy by Chu \& Kennicutt (1988), who observed high-velocity,
shock-accelerated material within the boundary of the SNR that is
distinct from the low-velocity quiescent material in the {\sc H~ii}
region. It was also identified in the radio by Ye et al. (1991), who
offer an explanation of why SNRs are expected to be found near the edge
of {\sc H~ii} regions based on the argument that sequential star
formation is expected to occur about 2.5 Myr after and 10 - 15 pc away
from the previous generations, as massive stars move away from their
birthplaces (Elmegreen \& Lada 1977).  UV analyses with the {\sl 
International Ultraviolet Explorer} (de Boer \& Savage 1980), {\sl HST}
STIS (Koenigsberger et al. 2000), and {\sl Far-Ultraviolet Spectroscopic
Explorer} ({\sl FUSE}; Hoopes et al. 2001) have also confirmed the
presence of an expanding structure. Danforth et al.  (2003) measured the
velocity of the SNR emission with {\sl FUSE}, and they found
high-velocity features with increasing velocity towards the center,
across the whole face of SNR B0057$-$724, {\sl typical for an expanding
shell}. The fastest emission ($v_{\rm LSR}\sim+335~{\rm km}~{\rm
s}^{-1}$) is found at about 4 pc south of HD~5980 and is consistent with
the peak in X-ray brightness.  From X-ray and radio images these authors
set the center of the SNR at 00$^{\rm h}$59$^{\rm m}$27$^{\rm s}$,
$-$72\deg10\arcmin15\arcsec (J2000).

Danforth et al. (2003) propose a topology, according to which a roughly
spherical SNR is located on the near side of N~66, basically closer to
us than NGC~346 itself. Consequently the rear side of the SNR is
propagating into relatively denser ionized nebula, while the near side
is propagating through the more diffuse SMC ISM, and is harder to
detect. The X-ray emission arises over a spherical cap, where the shock
has encountered and heated the denser material, for which Danforth et
al. (2003) estimate a pre-shock density of $\sim 6 \pm 2$~cm$^{-3}$.
According to these authors the remnant is not yet even halfway submerged
in the N~66 material, and thus, the observed radius of X-ray emission
(as observed by Naz\'{e} et al. 2002 and seen in the left panel of Fig.
\ref{fig-imaxoi}) is smaller than the actual blast wave radius. HD~5980
and other UV-bright stars in the NGC~346 association are embedded within
N~66, but HD~5980 lies behind the remnant, while NGC~346 is outside the
SNR\footnote{Danforth et al. suggest that Sk~80 is probably inside the
SNR.}. 


Danforth et al. (2003) also discuss the relationship between SNR
B0057$-$724 and N~66, and they conclude that the detection of {\sc O~vi}
emission together with the H$\alpha$ kinematics imply the presence of a
strong shock in N~66 associated with SNR B0057$-$724.  Although this
SN-driven shock affects its direct surroundings, it cannot serve as the
triggering mechanism of the observed star-forming event at the north of
the bar of N~66 for the following reason.  The expansion velocity of the
SNR is $v = 175$ km~s$^{-1}$ (Danforth et al.\ 2003).  Assuming the
Sedov solution for the SNR (e.g. Tang \& Wang 2005), the travel time
for the shock front to reach the observed star-forming region ($r \sim
18$ pc) is $t = 2r/5v = 0.04$ Myr, much shorter than the ages of the
compact PMS clusters located in the nebular arc, as estimated by us
(Paper~III; $\tau$~\lsim~2.5~Myr), and SSN07 ($\tau\approx3\pm1$~Myr).
Consequently, while the relative position of the SNR and the east-facing
feature seems to support a relation between them (Fig. \ref{fig-imaxoi};
left) SNR~B0057$-$724 {\sl cannot be} the triggering agent for the star
formation in the northern part of N~66. On the other hand, its massive
progenitor could have played a very important r\^{o}le to this event. 

\begin{figure*}[t!]
\epsscale{1.15}
\plotone{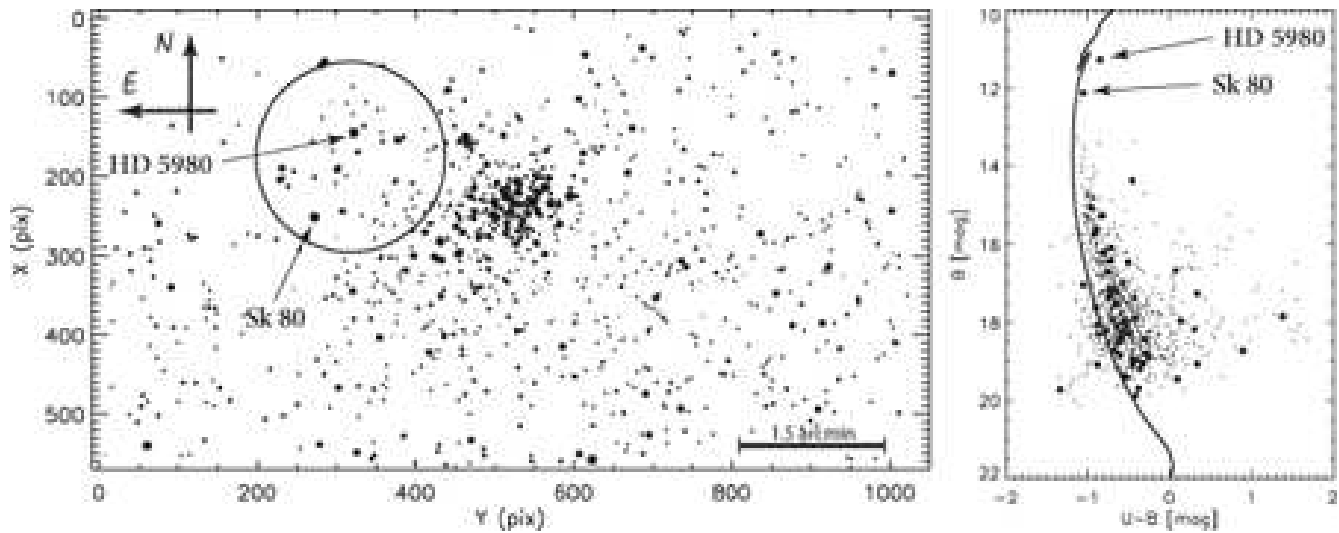}
\caption{Chart of all bright stars in the region of NGC~346/N~66 (left)
from photometry with the 0.9-m CTIO Telescope by Massey et al. (1989),
and the corresponding $U-B$, $B$ CMD (right).  The limits of the bubble
as it is outlined by the extended X-ray emission in the area within a
radius of $\sim$~18~pc is also plotted over the stellar chart. The stars
included by these limits are plotted with thick symbols in the CMD.  The
brightest sources in the region of the bubble, HD~5980 and Sk~80, are
indicated in both the chart and the CMD. A young isochrone of
$\log{\tau}\simeq 6.6$ (the youngest available) for the grid of
evolutionary models by Girardi et al. (2002) with no extinction applied 
is overplotted on the CMD for an assumed distance modulus of 
$m-M\simeq18.95$~mag.\label{fig-mapcmd}}
\end{figure*}

\section{A Star Formation scenario}

During their main sequence phase massive stars photo-ionize {\sc H~ii}
regions or blow bubbles through their winds. Shocks associated with the
ionization fronts or expanding bubbles can trigger cloud collapse, and
this process lasts for a few Myr, the lifetime of the star during the
main sequence phase. Such a process has been observed in the Milky Way,
e.g. at the periphery of the {\sc H~ii} regions Sh2-219 (Deharveng et
al. 2006) and RCW~79 (Zavagno et al. 2006). This mechanism may produce
young compact star clusters. Indeed, Lee \& Chen (2007) presented
recently their diagnosis of the r\^{o}le that massive stars play in the
formation of low- and intermediate-mass stars in the $\lambda$~Ori
region, Ori~OB1, and Lac~OB1 associations in the Milky Way. Their study
supports the radiation-driven implosion (e.g. Kessel-Deynet \& Burkert
2003) as the triggering mechanism, where the Lyman continuum photons
from a luminous O star create expanding ionization fronts to evaporate
and compress nearby clouds into bright-rimmed clouds. Implosive pressure
then causes dense clumps to collapse, prompting the formation of
intermediate-mass Herbig Ae/Be (HAeBe) stars somewhat deeper in the
cloud and low-mass (T~Tauri) PMS stars on the cloud surface (i.e., the
bright rim).

Bubbles are formed by the stellar winds, not the ionization energy of
massive stars. The expansion of an {\sc H~ii} region (not yet a bubble)
is caused by its higher thermal pressure due to the high temperature
compared to the pre-ionized state.  Typical expansion velocities are
around 10~km/s at temperatures $\sim 10^4$~K. On the other hand, the
expansion of a bubble is driven by the thermal pressure of the shocked
stellar wind in the central cavity. The velocity of this expansion
ranges from 10 to 100~km/s (Drake 1986), depending on the wind strength
and ambient density. When the massive star explodes as a supernova, it
explodes inside the bubble blown by itself during its main sequence
phase, and the produced SNR will have a size similar to the bubble blown
by the star. This will create the picture of a SNR with star formation
taking place on its edge, similar to what we observe at the northern
part of N~66. However, it is the massive progenitor that triggered star
formation and not the SN. The importance of stellar winds as a feedback
mechanism to star formation over supernovae has been demonstrated by
Leitherer et al. (1992), who present models for the integrated output of
mass, momentum, and energy from a concentration of massive stars,
computed for typical starburst parameters. In general, the most massive
stars evolve from the main sequence (MS) into blue supergiants (BSG) and
luminous blue variables (LBV) to become then Wolf-Rayet (WR) stars.
Leitherer et al. (1992) found that the (post-main sequence) OB phase,
together with the WR phase is the dominant contributor to the total wind
output. These authors considered the cases of an instantaneous and of a
continuous starburst and they found that in both cases the LBV phase is
unimportant, as a consequence of its short life time ($< 10^5$~yr; e.g.
Lamers 1989). A total power output between $10^{39.5}$ and
$10^{40.5}$~ergs~s$^{-1}$ is estimated by these authors for a
concentration of stars following a Salpeter IMF with masses between 1
and 120~M{\solar} throughout the epoch of a continuous starburst at
$10^{6} < T < 10^{7}$~yr. 



\subsection{Wind-Triggered Star Formation in N~66}

Considering the geometry of the north-northeastern part of the region of
NGC~346/N~66 as seen in Fig.~\ref{fig-imaxoi} ({\sl left}), the
hypothesis of a wind-triggered star formation event seems to explain
best the observed recent star formation in this area. We propose a
scenario according to which the massive progenitor of SNR~B0057$-$724
has triggered or contributed significantly to the triggering mechanism
of star formation through its expanding bubble, producing the northern
gaseous arc seen in [{\sc O~iii}] and H\alp. A schematic representation
of this scenario is shown in the right panel of Fig.~\ref{fig-imaxoi},
where the two important gaseous features, the southern part of the N~66
bar shaped by the photo-dissociation of the central OB stars and the
northern feature delineating the boundary of the SNR X-ray emission, are
drawn with red thick curves. If indeed the star-forming northern nebular
arc {\sl is} related to the activity at east, it would be interesting to
estimate how much mechanical energy could have been dumped in the area
to produce this event.


Naturally the existence of HD~5980 and Sk~80 in the same region with the
SNR may not be considered as a coincidence, and therefore one may
question if the progenitor is the only triggering agent or not. Indeed,
the massive object HD~5980 could contribute to the triggering process
through its own bubble blown during its main sequence phase, and another
hypothesis is that the cumulative action of even more massive stars in
the region, in addition to the SN-progenitor, may have triggered the
formation of the northern extended feature. This implies the existence
of a massive cluster in the area of the SNR as the triggering agent. In
order to test this hypothesis we used the most complete photometric
catalog of bright stars in the region (Massey et al. 1989) to search for
such a cluster. The stellar chart of all bright sources in this catalog
and the corresponding CMD, shown in Fig.~\ref{fig-mapcmd}, indicate that
no specific concentration of massive stars can be observed within the
bubble limits.  Support to the lack of a star cluster of any kind in the
region is also provided by the catalog of faint sources down to
$V\simeq$~28~mag from our ACS/WFC photometry of the whole NGC~346 region
(Paper~I), where it can be seen that no specific concentration of
low-mass stars exist in the area of the SNR either. It is most probable
that the bright stars seen in the CMD of Fig.~\ref{fig-mapcmd} belong to
the outskirts of NGC~346 itself or to the field. Nevertheless, the only
bright sources, which could provide the necessary kinetic energy for a
triggered star formation event, in addition to the SN-progenitor, are
HD~5980 (Sk~78) and Sk~80, both located within the SNR
(Fig.~\ref{fig-mapcmd}).

Single stars between 8 and 20~M{\solar} produce type II-P SNe when they
are red supergiants (Crowther \& Smartt 2007). Stellar structure models
of more massive stars predict that such stars will have final
carbon-oxygen cores of more than 3 M{\solar}, and hence form black holes
(Heger et al.  2003), suggesting that such objects form very faint SNe.
Recent discoveries suggest that very massive stars may explode as SNe in
their LBV stage. Pastorello et al. (2007) found that the very energetic
SN2006jc was a type Ic event, but embedded within a He-rich
circumstellar envelope and that it was coincident with an LBV-like
outburst just 2 yrs before collapsing. According to these authors it
could have been an originally very massive LBV which lost the last of
its H and He envelopes in energetic mass ejection episodes and collapsed
as a stripped Wolf-Rayet star. Recently Smith et al. (2007) reported the
analysis of the most luminous SN ever recorded, the type~IIn SN2006gy in
NGC 1260, which reached an absolute magnitude of $R \simeq -22$~mag. The
combination of optical and X-ray monitoring suggests it was the
explosion of a very massive star of about 100 M{\solar}, which failed to
shed its hydrogen envelope. According to these authors SN2006gy implies
that some of the most massive stars can explode prematurely during the
LBV phase, never becoming Wolf-Rayet stars.

Since there is no information about the nature of the progenitor of
SNR~B0057$-$727, and in order to have an upper-limit estimation of the
wind-blown energy expected by it, we may assume that this progenitor was
an object as massive as the estimate of Smith et al. (2007) for
SN2006gy, meaning roughly an early O-type BSG. Typical ionizing photon
fluxes for such stars are $F_{\rm Ly} \simeq
10^{50}$~UV~photons~s$^{-1}$ (Vacca et al.  1996; Schaerer \& de Koter
1997). O-type stars have mass-loss rates $\dot{M}$ of about
$10^{-7}$~-~$10^{-5}$~M{\solar}~yr$^{-1}$ and terminal velocities
$v_{\rm W}$ in the range 1~-~4~$\times 10^{3}$~km~s$^{-1}$ (Chlebowski
\& Garmany 1991; Lamers et al. 1995), thus injecting large amounts of
energy to the neighboring ISM. The corresponding mechanical luminosity
can be derived as $L_{\rm W} = \dot{M}v_{\rm W}^{2}/2$ (e.g. Cappa \&
Benaglia 1998). Consequently, an O3-type star with a mass loss rate of
$10^{-5}$~M{\solar}~yr$^{-1}$ and a wind velocity of 3750~km~s$^{-1}$
(Puls et al. 1996) would have a stellar wind mechanical luminosity of
$4.5 \times 10^{37}$~ergs~s$^{-1}$. This mechanical input is lower than
the assumed energy contribution of a massive cluster in the models by
Leitherer et al.  (1992), but since the density of the original cloud is
unknown, one can only speculate about the energy that would be needed to
trigger star formation in the region.  Nevertheless, the two additional
energy sources in the region should be also considered. Sk~80, being an
O7~I giant would contribute around $3\times 10^{36}$~ergs~s$^{-1}$ (Puls
et al. 1996). 

HD~5980 has been suggested to be a triple system (Koenigsberger et al.
2000) with the first component being a 50~M{\solar} WN6 star, the second
a 28~M{\solar} WN2 (with high uncertainty) and the third component being
a O4~-~O7 giant. Koenigsberger et al. (2000) found that a wind-wind
collision shock cone winds tightly around the second companion, and
using the relation given by Volk \& Kwok (1985), under a two-wind
approximation, they provide a rough estimate of $2 \times
10^{-4}$~M{\solar}~yr$^{-1}$ for the current mass-loss rate of the
system with a terminal velocity of $\sim$~2000~km~s$^{-1}$. This
mass-loss rate measurement, which is higher than typical rates of
Galactic WR stars (2~-~10~$\times 10^{-5}$~M{\solar}~yr$^{-1}$ with
terminal velocities of 1000~-~2500~km~s$^{-1}$; van der Hucht 1992),
yields a mechanical luminosity of $2.6\times 10^{38}$~ergs~s$^{-1}$ for
HD~5980. As far as the ionizing flux of HD~5980 is concerned, Law et al.
(2002) found that the Lyman continuum flux of WN stars appears to be
independent of spectral type, and lower bound estimates tend to cluster
around $10^{48}$~UV~photons~s$^{-1}$. This would be the ionizing energy
contributed by each of the WN components of HD~5980, while the third
component, being an O4~-~O7 giant, would provide an additional
$\sim$~$10^{49.5}$~UV~photons~s$^{-1}$.

From the mechanical luminosities of each of the involved objects in the
region, one can estimate that the total mechanical power that was
supposedly dumped in the region is of the order of that produced by
HD~5980 alone, $L_{\rm W}^{\rm Tot} \simeq 3\times
10^{38}$~ergs~s$^{-1}$, implying a small contribution from the massive
progenitor of SNR~B0057$-$724 and Sk~80 in the process. However, this
result stands only if the SN-progenitor did not go through a major
mass-loss event. According to the standard evolution of O-type stars
with masses \gsim~60~M{\solar}, these stars will go through a WNH or LBV
phase that removes via line-driven winds the remaining H-envelope to
yield a \lsim~20~M{\solar} WR star. Moreover, the large number of
observed giant eruptions of LBVs, which can remove around 10~M{\solar}
in a few years (Smith \& Owocki 2006), clearly suggests that massive
O-type stars do not shed most of their initial mass by the end of the
main sequence phase, but there is still enough mass remaining to power
major mass-loss during the WNH or LBV phase. Indeed, recent
observational evidence suggests that mass-loss rates for O stars need to
be revised downward (Smith 2007), in line with recent observed mass-loss
rates for clumped winds (e.g. Fullerton et al. 2006). 

Could the progenitor of SNR~B0057$-$724 have lost most of its initial
mass via an eruptive mass-loss event (during e.g. a LBV phase),
triggering the observed star formation? The answer is positive. The
primary star in the $\eta$~Car system, which is suggested to have a
present-day mass \gsim~100~M{\solar} has lost 20~-~30~M{\solar}, in
violent LBV eruptions in just a few thousand years (Smith et al. 2003),
in addition to its steady wind. Consequently a SN-progenitor alone is
capable of triggering star formation event, as has been recently
observed with the {\sl AKARI} infrared satellite by Koo et al. (2008).
These authors report the discovery of a star-forming loop around the
young, Crab-like SNR~G54.1+0.3 in the Galaxy, and they propose that star
formation was triggered by the progenitor star of G54.1+0.3, which has a
mass of \lsim\ 15 M{\solar}. The triggering must have occurred near the
end of the progenitor's life, possibly after it had evolved off the main
sequence. Under these circumstances, the massive progenitor of
SNR~B0057$-$724 alone may have provided the necessary mechanical energy
to trigger star formation to the north, while the contribution of
HD~5980 may have significantly enhanced this event.


\begin{figure*}[t!]
\epsscale{1.}
\plotone{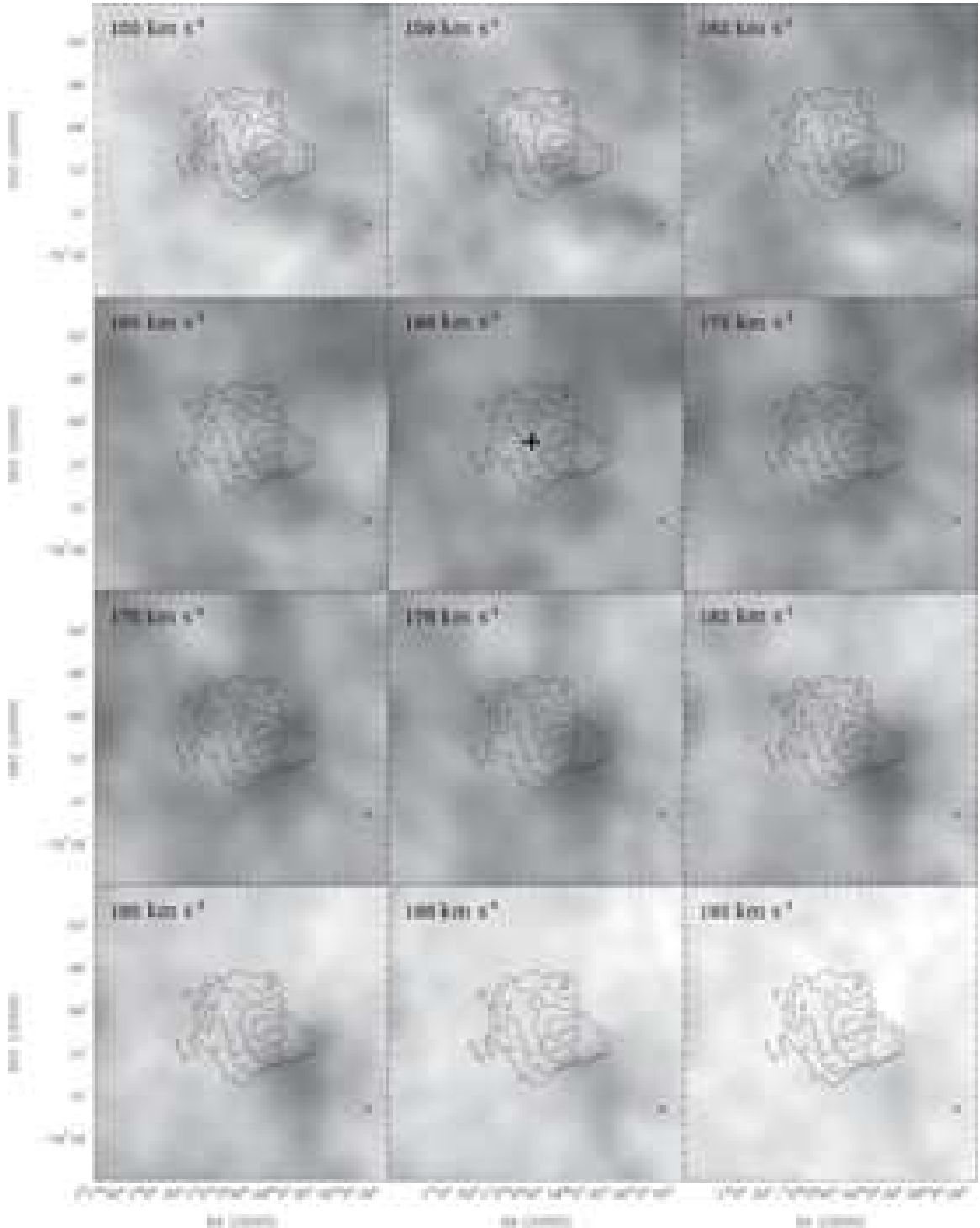}
\caption{Images of the region of NGC~346/N~66 from the {\sc H~i} mosaic
constructed with {\sl ATCA} and {\sl Parkes} (greyscale) with the H$\alpha +$
continuum from MCELS overlaid (contours). The most significant velocity
planes are shown, with the corresponding heliocentric velocity indicated
on the top-left corner. Each image covers an area about
0.35\deg~$\times$~0.35\deg\ ($\sim 357 \times 357$~pc$^{2}$) wide,
centered on N~66, which is indicated by the H\alp\ isopleths. North is
up, East is to the left. The thick cross-symbol in the image of the
168~km~s$^{-1}$ velocity plane indicates the position of
SNR~B0057$-$724. A hole in the {\sc H~i} east of N~66 and possibly
related to the area of the SNR can be seen in the velocity planes
between 165 and 178~km~s$^{-1}$, with its larger extend seen in
the 168~km~s$^{-1}$ velocity plane.  This velocity coincides almost
perfectly with the systematic velocity of the SNR found with echelle
spectroscopy by Chu \& Kennicutt (1988), indicating that the neutral
hydrogen {\sl is possibly} kinematically related to the ionized
gas.\label{fig-hi}}
\end{figure*}

\subsection{Gas Kinematics in the Area of the SNR}

Additional information on the interstellar environment of
SNR~B0057$-$724 can certainly be provided by the kinematic study of the
surrounding gas. In Fig. \ref{fig-hi} we show a set of images centered
on NGC~346/N~66, constructed from observations of atomic and ionized
gas. The observations from the neutral hydrogen emission survey of the
SMC constructed with the {\sl Australia Telescope Compact Array} (ATCA)
and {\sl Parkes Observatory} at the {\sl Australia Telescope National
Facility} (ATNF) are shown as greyscale images\footnote{{\sl Parkes}
single-dish observations were combined with an aperture synthesis mosaic
of 320 separate pointings of the 375-m array from {\sl ATCA}
interferometry. These observations are sensitive to angular scales
between 98{\arcsec} (30~pc) and 4{\deg} (4~kpc) over a field of 20
deg$^2$ with velocity channels $\sim$1.65~km~s$^{-1}$ wide, spanning
between heliocentric velocities of $\sim$88.5 and
$\sim$215.5~km~s$^{-1}$.} (Staveley-Smith et al. 1997; Stanimirovi\'{c}
et al. 1999). The original data-cubes were kindly provided by Snezana
Stanimirovi\'{c}. Overplotted are the contours from the H\alp\ image of
the same general area of N~66 obtained within the {\sl Magellanic Cloud
Emission Line Survey}\footnote{MCELS is accessible at {\tt
http://www.ctio.noao.edu/mcels/}} (MCELS; Smith et al. 2000). Only
isopleths, which correspond to intensity higher than 3$\sigma$ above the
background, where $\sigma$ is the standard deviation of the background
noise, are shown. 



In Fig. \ref{fig-hi} the 12 most significant velocity planes from the HI
interferometric data between 155 and 192 km~s$^{-1}$ (heliocentric
velocities) are shown. Each plane is 3.3~km~s$^{-1}$ wide and it is the
average of two planes in the processed cube. The central compact H\alp\
concentration in each image is the nebula N~66 and the area of the SNR
is directly to its left (east). It is very interesting to see that from
these sequential images, indeed, {\sl a possible relation between the
gas kinematics and the area of SNR~B0057$-$724} is revealed through {\sl 
a gas bubble visible in the planes between 165 and 178 km~s$^{-1}$}. 
However, the hole seen in {\sc H~i} seems to be much larger than the SNR
itself, which is about 2\arcmin\ in diameter (Chu \& Kennicutt 1988).
This implies that this hole should have been blown also by HD~5980 and
perhaps with the help of other stars. It should also be mentioned that
the location of the SNR is at the edge of the {\sc H~i} hole, not its
center. From the quantitative point-of-view, the echelle data of Chu \&
Kennicutt (1988) indicate a systemic velocity for the SNR of $v_{\rm
LSR} \sim$~158~km~s$^{-1}$, which translates to heliocentric velocity of
$v \approx v_{\rm LSR} + 11~{\rm km~s}^{-1} \simeq 169$~km~s$^{-1}$.
This value {\sl is in very good agreement with the velocity plane where
the hole in the atomic gas at the area of the SNR is more clearly
visible} in the images of Fig. \ref{fig-hi} (velocity of
168~km~s$^{-1}$). This result provides evidence that indeed the atomic
gas {\sl is possibly} kinematically related to the ionized gas and
therefore possibly to the SNR. 

\begin{figure}[t!]
\epsscale{1.15} 
\plotone{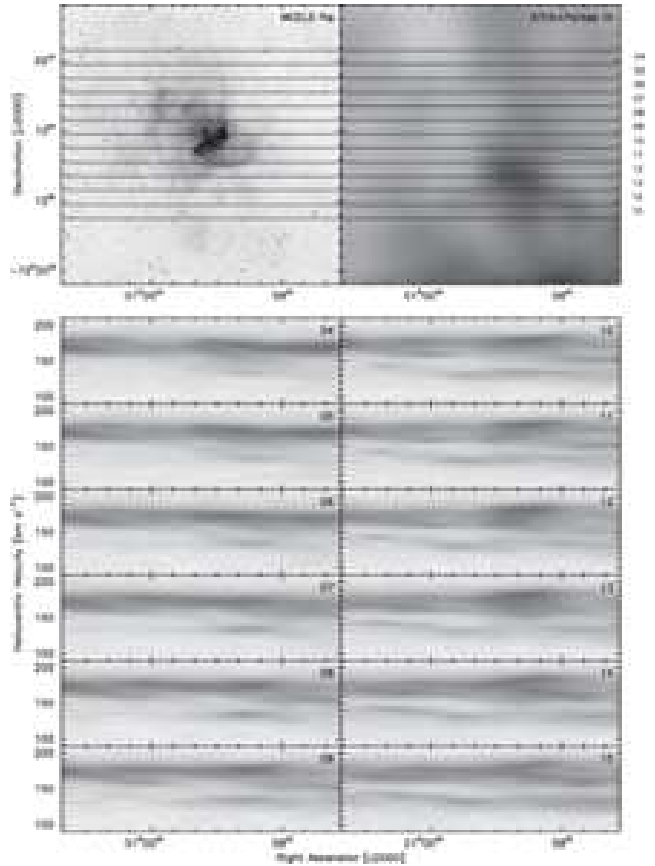} 
\caption{Position-velocity diagrams from ATCA and Parkes HI observations
for 12 selected slits covering the whole area of N~66 from east to west
(long-slit spectra). The positions of the slits are shown on the maps of
the area in H{\alp} (from MCELS) and HI (from ATCA and Parkes). The HI
velocity of $\sim$~170~km~s$^{-1}$ shows a split at slit position 08 and
southwards, indicating that the shell probably expands. This expanding
shell structure continues south.\label{fig-hiewcuts}}
\end{figure}

However, it is not clear whether the {\sc H~i} hole, seen at
168~km~s$^{-1}$, is indeed a bubble. In order to clarify if this hole
actually represents an {\sc H~i} bubble, we examine the spatio-kinematic
structure of the neutral hydrogen in the vicinity of N~66 by
constructing position-velocity (PV) diagrams. Each PV diagram consists
of a east-west slice (slit) through the {\sc H~i} data cube where we
have averaged over 1\arcmin\ in the north-south direction. Multiple PV
diagrams were constructed, each after stepping sequentially by 1\arcmin\
in the north-south direction to cover the whole area of N~66 from east
to west. Twelve PV diagrams are shown in Fig.~\ref{fig-hiewcuts}, where
the positions of the corresponding slits are overlayed on the maps of
the area in H{\alp} (from MCELS) and {\sc H~i} (from ATCA and Parkes).
The vicinity of N~66 is fully covered in the H{\alp} image, the features
of which are easily comparable to the color-composite image of
Fig.~\ref{fig-imaxoi}. The star-forming feature north of the bar of N~66
is also easily detectable in this map. Above this feature, at slit
position 07, the {\sc H~i} velocity does not show any split. At slit
position 08, though, the {\sc H~i} at $\sim$~170~km~s$^{-1}$ starts to
split, and further south, this velocity split becomes more clear, giving
possible evidence of an expanding shell-structure. The PV diagrams of
Fig.~\ref{fig-hiewcuts} indicate an expansion velocity of
$\sim$~10~km~s$^{-1}$. It is interesting to note that the line split
does not occur over N~66, at slit positions 08 and 09. This indicates
that N~66 does not seem to be responsible for the expanding shell.

Considering a distance of 60~kpc and a correction factor for He of 1.4
we integrated the {\sc H~i} column density over an area roughly
12\arcmin~$\times$~12\arcmin wide, centered on the shell, between
velocities of 158 and 188 km~s$^{-1}$ and derived a total mass of
1.52~$\times$~10$^6$~M{\solar} in the {\sc H~i} expanding shell. This
mass corresponds to a total kinetic energy of about
1.5~$\times$~10$^{51}$~erg. Taking into account the total mechanical
power estimated in \S~3.1 for the SN-progenitor, HD~5980 and Sk~80
(equal to 3~$\times$~10$^{38}$~erg~s$^{-1}$), one can see that these
objects have dumped a larger amount of mechanical energy than the
estimated kinetic energy in the shell in less than 1~Myr. Indeed, it
seems that it is typical for expanding bubbles that the kinetic energy
retained in the shell is only a small fraction of the total stellar
energy injected. This is observed for example in the superbubble N~51D
in the Large Magellanic Cloud, where the {\sc H~i} kinetic energy is
found to be almost an order of a magnitude lower that the total stellar
wind and supernova energy input (Chu et al. 2005). Based on our
calculations of \S~3.1, the total energy provided by the SN-progenitor,
HD~5980 and Sk~80 for the last couple Myr is indeed an order of a
magnitude higher than the kinetic energy in the shell. 





\section{Final Remarks}


In conclusion, based on the evidence presented here, we propose an
expanding {\sc H~ii} region or bubble blown by the winds of the massive
progenitor of SNR~B0057$-$724, possibly the massive object HD~5980 and
maybe the O7 giant Sk~80 - all located at the eastern vicinity of
NGC~346/N~66, as the secondary mechanism that shapes the recent star
formation in this region, in addition to the photo-ionizing process of
the OB stars of the association. This mechanism is similar to shell-like
{\sc H~ii} regions, with a central young cluster in an evacuated cavity,
and with ongoing star formation triggered around their periphery, and
therefore is a rather quiescent process. Typical examples for such
regions in the MCs are LH~9/N~11 (Walborn \& Parker 1992) and 30 Doradus
(Walborn et al. 2002) in the LMC and NGC~602/N~90 (Gouliermis et al.
2007; Carlson et al. 2007) in the SMC. It should be noted, however, that
within the X-ray bubble associated with the SNR, which most probably
outlines the wind-blown bubble of its progenitor, no signature of any
underlying cluster is found.

The induced star formation in this part of the region of NGC~346/N~66 is
evidenced by the existence of PMS clusters, candidate YSOs and
IR-emission peaks at a projected distance $\sim$ 18 pc and possibly at
even larger distances from the X-ray bubble. All these objects are
preferably located in an extended feature north of the bar of N~66,
which is outlined by the emission from the gas, observed in H\alp\ and
[{\sc O~iii}], as well as from the dust, as it is seen in CO, H$_2$ and
8\micron\ maps. Lee \& Chen (2007) findings in the $\lambda$~Ori region,
Ori~OB1, and Lac~OB1 associations suggest that intermediate-mass HAeBe
stars are formed in the denser parts of the cloud, while the low-mass
T~Tauri stars are formed at the outskirts near the photo-evaporating
cloud layers. Our photometry in H{\alp} (Paper~III) showed that indeed
stars with H{\alp}-excess (signature of candidate HAeBe stars) are
located in the area of the northern compact PMS clusters. Based on the
aforementioned observations in different wavelengths, the extended
feature north of the bar of N~66, which hosts this star formation, may
be considered as part of a large arc centered on the SNR
(Fig.~\ref{fig-imaxoi} {\sl right}), which extends also to the south and
in the bar. In this case, one may conclude that its northern part is
projected closer to us, outside N~66, since the dust is well outlined by
the 8\micron\ emission (seen in red in Fig.  \ref{fig-imaxoi}; left),
while its southern part is embedded {\sl in} the nebula (as suggested by
the schematic of Fig.~\ref{fig-imaxoi}; right). 

On the other hand, considering that the northern part of this feature as
seen in CO and IR maps appears to be more linear with small curvature
(see e.g. maps by Contursi et al. 2000; Rubio et al. 2000; Simon et al.
2007), one may argue that it may not bear a morphological connection to
the triggering sources to the east. Against this argument stands the
fact that this linearity is confined only in the northernmost part of
the structure, while its southern part shows a curvature to the east,
and fits very well to the alignment of emission from both gas and dust
towards the south (see e.g. maps by Contursi et al. 2000; Rubio et al.
2000; Simon et al. 2007; Paper~III). The latter appears to be well
related to the positions of the {\sc H~ii} regions and IR peaks, which
are located {\sl in} the south-eastern part of the bar of N~66.
Consequently, the northern feature can be considered to be extended to
the south and to become {\sl embedded} in the south-eastern part of the
bar, forming a large arc-like feature, which covers the whole eastern
part of N~66 nebula (Fig.\ref{fig-imaxoi}), and which is centered on the
SNR. Further support to this argument provides the outstanding
coincidence of this arc structure and the non-thermal emission related
to the SNR~B0057$-$724, shown in the H\alp\ and {\sc [S~ii]}
continuum-subtracted images by Reid et al. (2006; their figures 1 and
2). Under these circumstances the dust emission from the northernmost
linear part of the structure may not come from only a ``strip'' of
molecular material, but from a whole curved surface of the wind-blown
bubble, projected in front of the bar of N~66. It should be noted,
though, that an equally valid description of the morphology of the
8\micron\ emission is that of a partial arc with mild curvature pointing
west, which appears connected to the N~66 bar through a clump of
emission.

In any case, the proposed scenario draws the picture of a unique case of
complexity in the star formation history of NGC~346/N~66, the brightest
{\sc H~ii} region in the SMC. According to this scenario it is the
collaborative (or competitive) action of two major energy sources that
shapes the current star formation process in this extraordinary region.
The first source, located at the center of the nebula N~66, is the
bright stellar content of the association NGC~346 and affects mostly the
bar of the nebula. The second source, located at the eastern part of the
region, is associated with the bright X-ray bubble of a SNR. We propose
that the massive progenitor and possibly two other massive stellar
objects have triggered the most recent star formation at the northern
part of the region away from the bar of N~66 within that last few Myr. 
Observations of the gas kinematics provide further support to this
hypothesis. Under these circumstances, the identified compact PMS
sub-clusters in the region {\sl cannot} all be the product of
spontaneous cloud collapse and fragmentation (see e.g. Elmegreen \& Lada
1977). They are rather the product of star formation inside {\sl and}
outside the bar of N~66, induced by wind-shock waves (e.g. Vanhala \&
Cameron 1998) or ionization shock fronts (e.g. Kessel-Deynet \& Burkert
2003) driven by OB stars into the surrounding cloud.


\acknowledgements We wish to thank Lars Christensen (ESA) for producing
the beautiful color-composite image of NGC~346 presented in Figure
\ref{fig-imaxoi} (left). We are also grateful to Snezana
Stanimirovi\'{c} for providing the original {\sl ATCA} and {\sl Parkes}
combined {\sc H~i} data-cubes of the SMC. D.A. Gouliermis kindly
acknowledges the support of the German Research Foundation (Deutsche
Forschungsgemeinschaft - DFG) through the individual grant GO 1659/1-1.
Based on observations made with ESO Telescopes at the La Silla
Observatories under program ID 56.C-0379, with XMM-Newton, an ESA
science mission with instruments and contributions directly funded by
ESA Member States and NASA, and with the Spitzer Space Telescope, which
is operated by the Jet Propulsion Laboratory, California Institute of
Technology under a contract with NASA. The Australia Telescope Compact
Array and the Parkes Telescope are part of the Australia Telescope,
which is funded by the Commonwealth of Australia for operation as a
National Facility managed by CSIRO.



\end{document}